\documentclass[12pt,preprint]{aastex}
\usepackage{rotating}
\usepackage{epsfig}
\usepackage{ifpdf}

\begin{document}

\title{JVLA Observations of Young Brown Dwarfs}

\author{Luis F. Rodr\'\i guez\altaffilmark{1}, Luis A. Zapata\altaffilmark{1} and Aina Palau\altaffilmark{1}}

\altaffiltext{1}{Instituto de Radioastronom\'\i a y Astrof\'\i sica, 
UNAM, Apdo. Postal 3-72 (Xangari), 58089 Morelia, Michoac\'an, M\'exico}

\email{l.rodriguez,l.zapata,a.palau@crya.unam.mx}
 
\begin{abstract}
We present sensitive 3.0 cm JVLA radio continuum observations of six regions of low-mass star
formation that include twelve young brown dwarfs and four young brown dwarf candidates.
We detect a total of 49 compact radio sources in the fields observed, of
which 24 have no reported counterparts and are considered new detections. Twelve of the radio sources show variability in timescales of weeks to months, suggesting
gyrosynchrotron emission
produced in active magnetospheres. Only one of the target brown dwarfs, FU Tau A, was detected. However, we detected radio emission associated with two of the
brown dwarf candidates, WL 20S and CHLT 2. The radio flux densities of the sources associated with these brown dwarf candidates are more than an order of magnitude larger than
expected for a brown dwarf and suggest a revision of their classification. In contrast, FU Tau A falls on the well-known correlation between radio luminosity and bolometric luminosity, suggesting that the 
emission comes from a thermal jet and that this brown dwarf seems to be forming as a scaled-down version of low-mass stars.

\end{abstract}  


\section{Introduction}

Brown dwarfs (BDs) are intriguing objects whose masses lie \rm in between the mass of planets and the mass of stars. They are supposed to burn 
deuterium in their interiors, thus having a mass larger than 13 $M_{Jup}$, but their mass is not large enough to drive stable hydrogen burning (where stable means sufficient to 
maintain thermodynamic equilibrium), corresponding to masses smaller 
than about 75 $M_{Jup}$. The main difference between BDs and stars is that BDs are not supported by thermal pressure, as they have not sufficient internal heating\rm, but are 
supported instead by electron degeneracy pressure. Thus, from the point of view of their internal structure, BDs are more similar to giant planets than to stars. However, there is currently a 
debate about how to define the borderline separating BDs and planets. While the traditional view is the aforementioned condition of deuterium burning for BDs 
(IAU definition 2003\footnote{http://home.dtm.ciw.edu/users/boss/definition.html}), this has 
been questioned by several authors such as Chabrier et al. (2007, 2014). These authors present counterexamples which do not fulfill the deuterium-burning criterion, and propose 
that a better criterion could be the formation mechanism: while BDs and stars are most likely formed through collapse of a protostellar core (as opposed to a disk), planets are most likely formed
through the `core accretion' model in a protoplanetary disk (Pollack et al. 1996).

Therefore, it is crucial to test from a solid observational base that BDs form indeed in a similar way as low-mass stars. Recently, a number of statistical studies of 
BDs and young stellar objects (YSOs) in the Class II/III stages (according to the evolutionary scheme of Adams et al. 1987) seem to indicate that the formation mechanism of 
BDs cannot be easily distinguished from the formation of stars (e.g., Bayo et al. 2011; Scholz et al. 2012a; Luhman 2012; Alves de Oliveira et al. 2013; Downes et al. 2014). 
Most of these studies aim at sampling the IMF down to the substellar regime, or studying the spatial distributions of stars and BDs. Other studies have compared accretion properties 
of low-mass YSOs with those of BDs, finding that they are consistent with a common formation mechanism (e.g., Muzerolle et al. 2005, Downes et al. 2008, Alcal\'a et al. 2014). 
The presence of disks in stars as well as in BDs also point to a common formation mechanism (e.g. Scholz et al. 2006; Ricci et al. 2014). An additional 
and crucial complementary study to test if BDs form as low-mass stars should come from studying their centimeter emission, as YSOs are known to emit at these wavelengths 
either through free-free emission from thermal radio jets when they are very young or through gyrosynchrotron emission from active magnetospheres later in their evolution (e.g. Feigelson \& Montmerle 1999;
Dzib et al. 2015).

The YSOs emitting gyrosynchrotron emission are typically in the Class II or later stages, and their emission is highly time variable (e.g., Dzib et al. 2013, 2015; Pech et al. 2016). 
On the other hand, thermal radio jets are non-variable and associated typically with more embedded YSOs (in the Class 0/I stage, e.g., Beltr\'an et al. 2001, Ward-Thompson et al. 2011; 
Carrasco-Gonz\'alez et al. 2012). In particular, this last possibility is especially compelling because signs of outflows have been clearly found in BDs, both as optical jets 
(e.g., Whelan et al. 2009, 2014; Joergens et al. 2012) and as molecular outflows (e.g., Phan-Bao et al. 2011, 2014; Monin et al. 2013; Palau et al. 2014). However, BDs have rarely been detected at centimeter wavelengths (e.g.,
Krishnamurthi et al. 1999;  G\"udel 2002; Osten \& Jayawardhana 2006), but this could be due to a lack of sensitivity of the instruments used in earlier work. Actually, recent studies have shown that the 
new capabilities of the VLA allow the detection of very faint sources which are associated with proto-BD candidates (Morata et al. 2015), thus being candidates to be thermal radio jets driven by substellar objects.

In this paper, we present deep observations 
with the Jansky VLA toward six BDs associated with known indicators of outflows, and with no published detection in the radio continuum. Our main aim is to study the nature of their 
centimeter emission and to compare to the emission typically found in YSOs. In Section 2 we describe the observations. In Section 3 we present the detections, and in Section 4 we 
discuss our detections in the context of the well-known properties of centimeter emission from YSOs.

%(We will assume that brown dwarfs are objects with a spectral class of M6.5 or later.)

\section{Observations}
The observations were made as part of project 14B-230 with the Karl G. Jansky Very Large Array of NRAO\footnote{The National 
Radio Astronomy Observatory is a facility of the National Science Foundation operated
under cooperative agreement by Associated Universities, Inc.} centered at the rest frequency of 9.9 GHz
(3.0 cm) during
2014 October, November and December.  At that time the array was in its C configuration providing a maximum baseline of 3.4 km and an
angular resolution of $\sim2''$ at the wavelength of 3.0 cm. The field of view is taken to be the full width at half power of the primary beam ($4\rlap.'2$ at
the observing wavelength), although in the case of sources that are bright enough, imaging can be made outside of this region. The phase centers of the six regions observed
are given in Table 1. In this Table we also list the number of times each region was observed (from 3 to 5 times), the synthesized beam and the rms of the final
image and the amplitude and gain calibrators.

The digital correlator of the JVLA was configured in 32 spectral windows of 128 MHz width divided 
in 64 channels with spectral resolution of 2 MHz each.
The total bandwidth was about 4 GHz in a dual-polarization mode.
The half power full width of the primary beam is $4\rlap.'6$ at 3.0 cm. 

The data were analyzed in the standard manner using the CASA (Common Astronomy Software Applications) package of NRAO,
although for the stages of component fitting and image contouring we used the AIPS (Astronomical Image Processing System)
package. 
We used the ROBUST parameter of CLEAN set to 2, to obtain a better sensitivity at the expense of losing some 
angular resolution. In Table 2 we list the sources detected with their counterparts when these exist, the positions, total flux densities and notes on the time variability,
angular extent and nature of the sources. The counterparts were searched using the SIMBAD (The Set of Identifications, Measurements, and Bibliography for Astronomical Data)
database.
The 12 sources that have significant variability over the period of the observations exhibit maximum variations (taken to be the ratio between the maximum
and minimum flux densities observed considering all epochs) that go from factors of 1.5 to 3.7.

\begin{figure}
\centering
\vspace{-2.8cm}
\includegraphics[angle=0,scale=0.8]{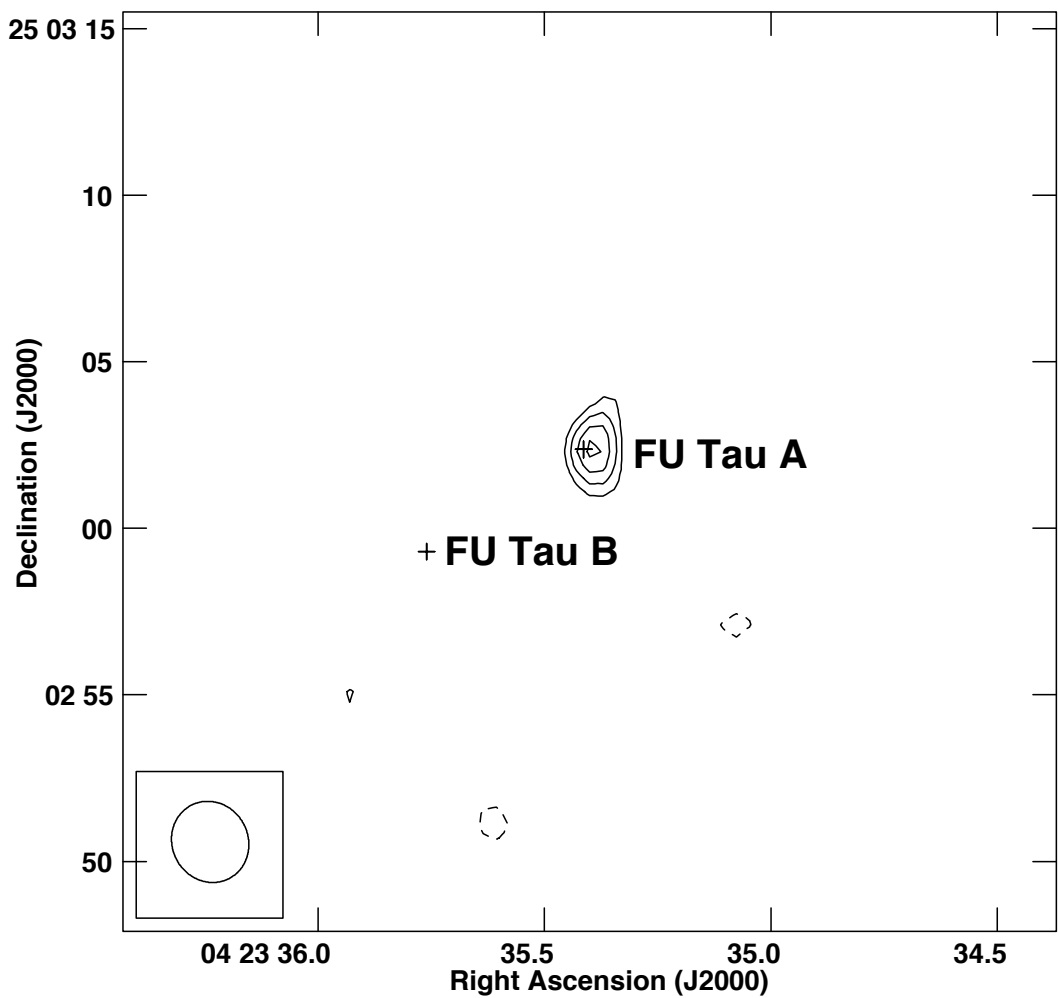}
\vskip-2.0cm
\caption{\small JVLA 3.0 cm continuum contour image of the FU Tau region.
The contours are -4, -3, 3, 4, 5, and 6 
times 1.55 $\mu$Jy beam$^{-1}$, the rms of the image. The negative contours are shown as dashed lines.
In the following contour images the negative contours are listed in the caption even if there are no such values
in the image.
The half-power contour of the synthesized beam of the image is shown in the bottom left corner. The crosses mark the positions
of the brown dwarfs FU Tau A and FU Tau B from Cutri et al. (2003), corrected for proper motions as described in the text.}
\label{fig1}
\end{figure}

\pagebreak

\begin{figure}
\centering
\vspace{-1.2cm}
\includegraphics[angle=0,scale=0.8]{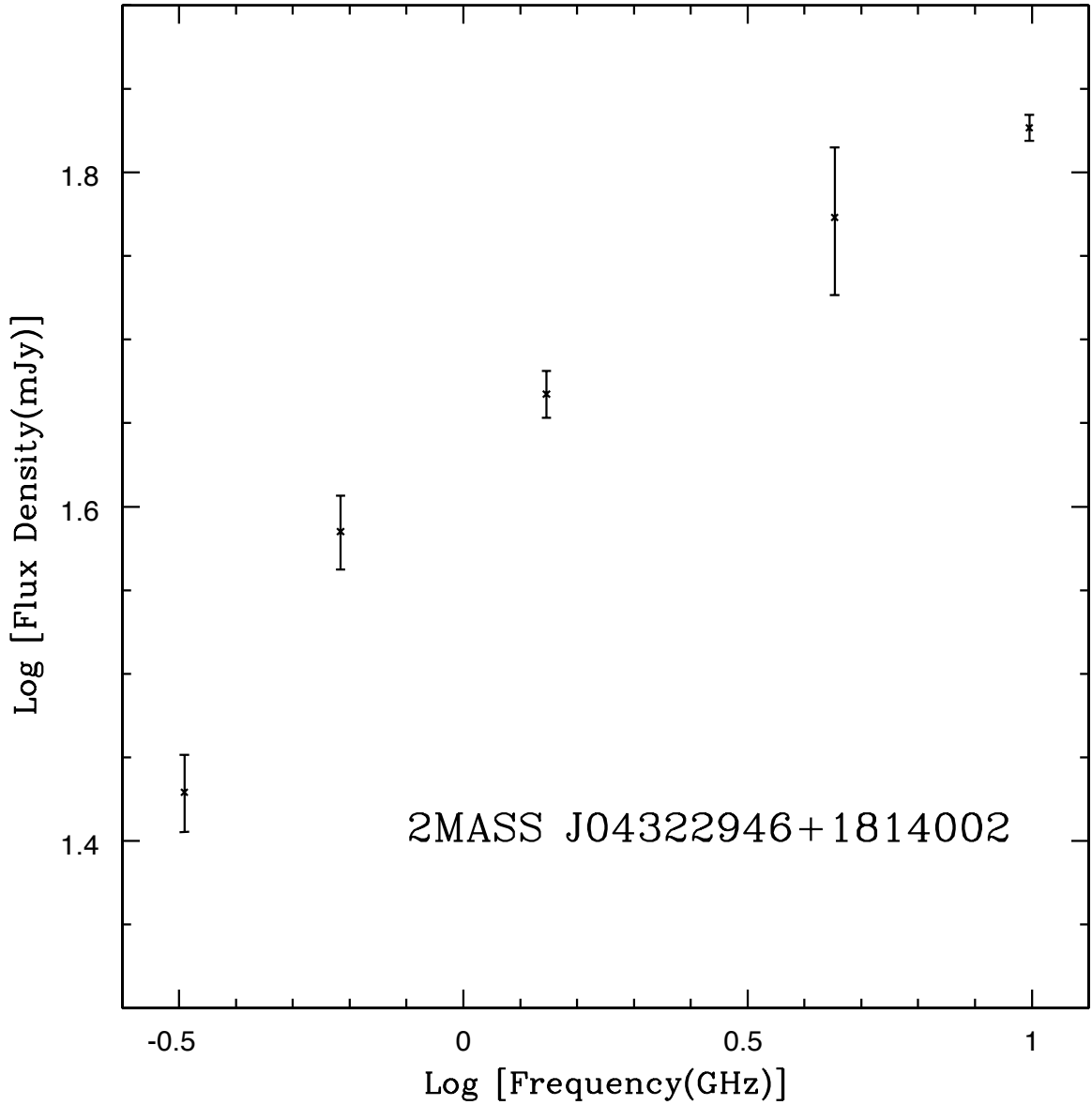}
\vskip-1.5cm
\caption{\small  Radio spectrum of the source 2MASS J04322946+1814002, located in the MHO 5 region. 
The data points at 323 and  608 MHz are from Ainsworth et al. (2016),while the data points at 1.4, 4,5 and 9.9 GHz are 
from Condon et al. (1998), Dzib et al. (2015) and this paper, respectively.
}
\label{fig2}
\end{figure}
 
\pagebreak

\begin{figure}
\centering
\vspace{-2.8cm}
\includegraphics[angle=0,scale=0.8]{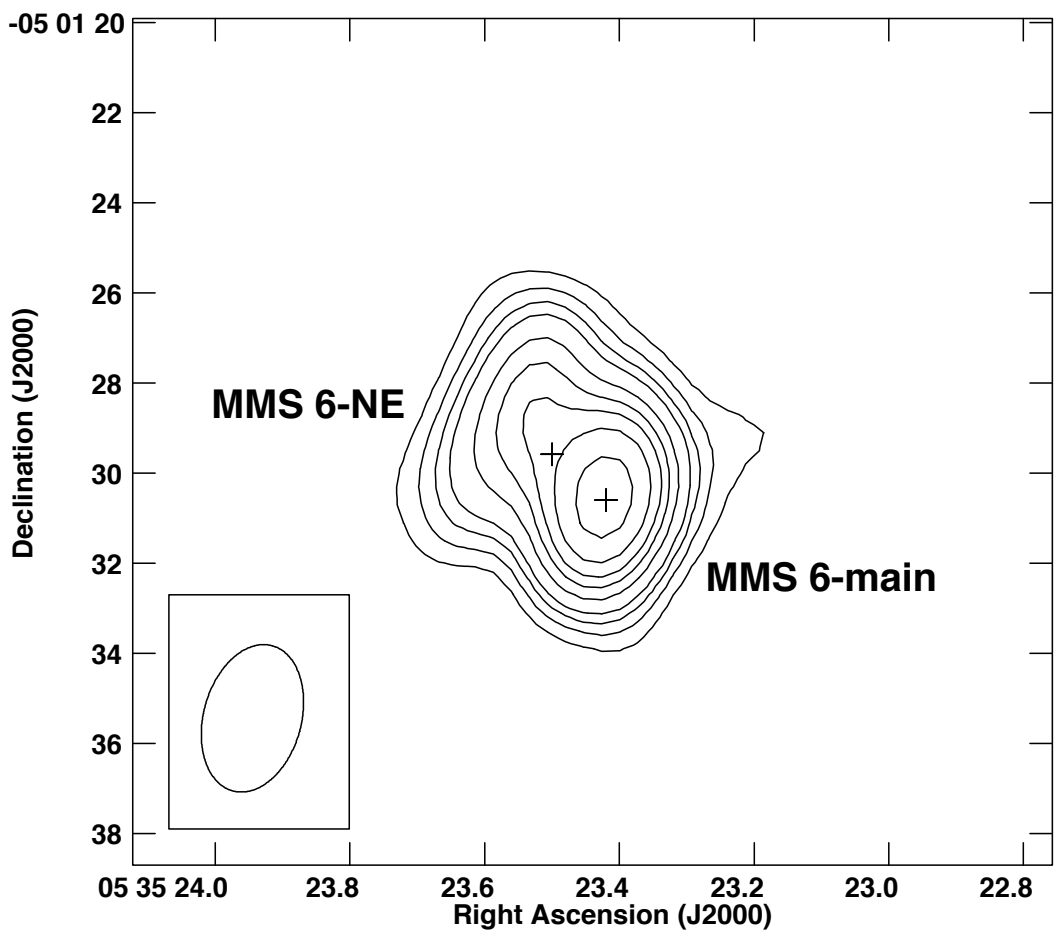}
\vskip-2.0cm
\caption{\small JVLA 3.0 cm continuum contour image of the MMS 6 region.
The contours are -4, -3, 3, 4, 5, 6, 8,  10, 12, 15 and 20
times 4.60 $\mu$Jy beam$^{-1}$, the rms of the image.
The half-power contour of the synthesized beam of the image is shown in the bottom left corner. The crosses mark the positions
of the mm sources MMS 6-main and MMS 6-NE from Takahashi et al. (2009).}
\label{fig3}
\end{figure}

\pagebreak

\begin{figure}
\centering
\vspace{-2.8cm}
\includegraphics[angle=0,scale=0.8]{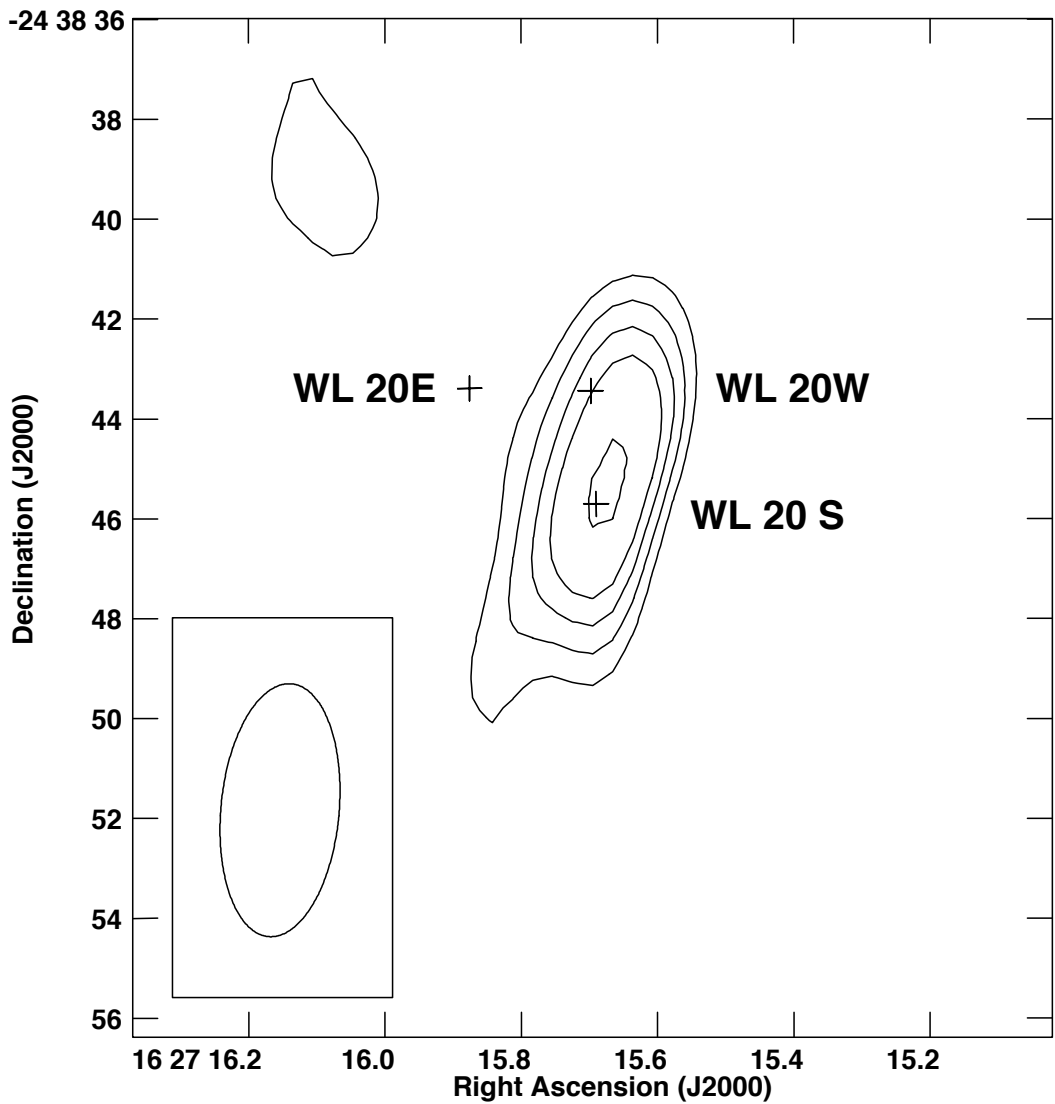}
\vskip-2.0cm
\caption{\small JVLA 3.0 cm continuum contour image of the WL 20 region.
The contours are -4, -3, 3, 4, 5, 6, and 8
times 29 $\mu$Jy beam$^{-1}$, the rms noise in this region of the image.
The half-power contour of the synthesized beam of the image is shown in the bottom left corner. The crosses mark the positions
of the stars WL 20E, WL 20W and WL 20S from Cutri et al. (2003), Alves de Oliveira (2010) and Barsony et al (2012). The radio emission is associated with WL 20S, the brown dwarf candidate.}
\label{fig4}
\end{figure}

\pagebreak

\begin{figure}
\centering
\vspace{-2.8cm}
\includegraphics[angle=0,scale=0.8]{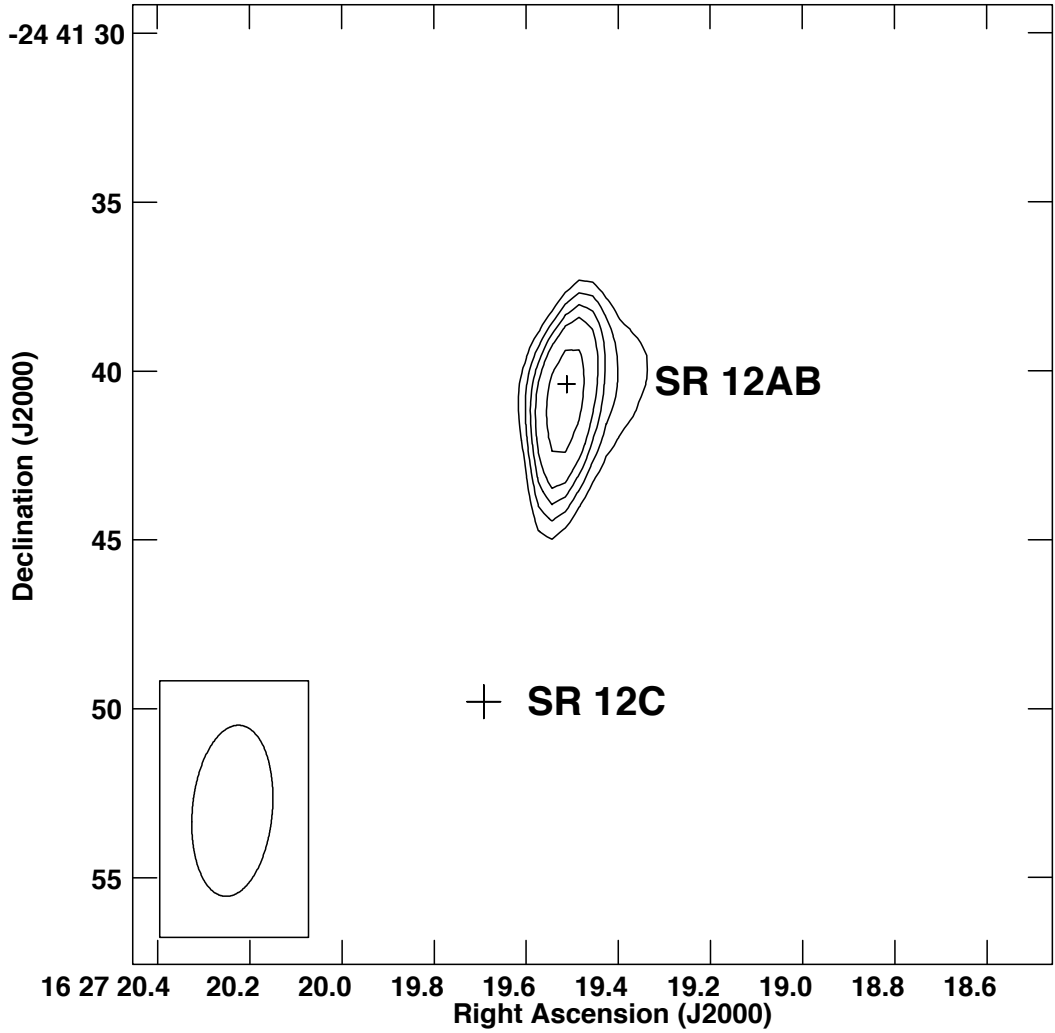}
\vskip-2.0cm
\caption{\small JVLA 3.0 cm continuum contour image of the SR 12 region.
The contours are -4, -3, 3, 4, 5, 6, and 8
times 8 $\mu$Jy beam$^{-1}$, the rms noise in this region of the image.
The half-power contour of the synthesized beam of the image is shown in the bottom left corner. The small cross marks the position of the T Tau binary SR 12AB from Cutri et al. (2003)
and the large cross marks the position of the candidate planet SR 12C, as estimated from the image of Kuzuhara et al. (2011).}
\label{fig5}
\end{figure}

\pagebreak

\begin{figure}
\centering
\vspace{-2.8cm}
\includegraphics[angle=0,scale=0.8]{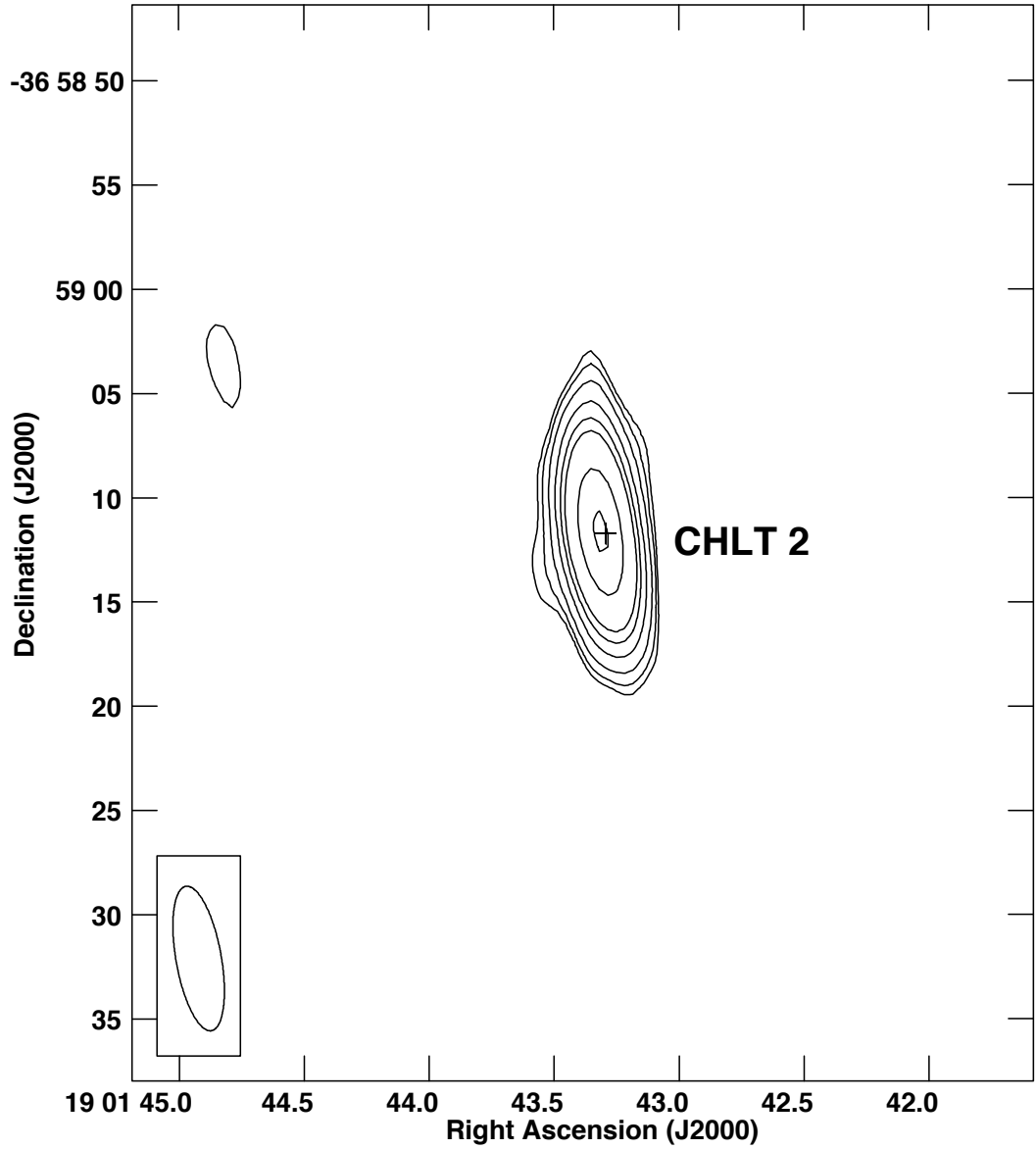}
\vskip-2.0cm
\caption{\small JVLA 3.0 cm continuum contour image of the CHLT 2 region.
The contours are -4, -3, 3, 4, 6, 10, 15, 20, 40 and 60
times 8 $\mu$Jy beam$^{-1}$, the rms noise in this region of the image.
The half-power contour of the synthesized beam of the image is shown in the bottom left corner. The cross marks the position of CHLT 2 from the near-infrared survey of
Haas et al. (2008; their source 1005).}
\label{fig6}
\end{figure}

\pagebreak

\begin{figure}
\centering
\vspace{-2.8cm}
\includegraphics[angle=0,scale=0.6]{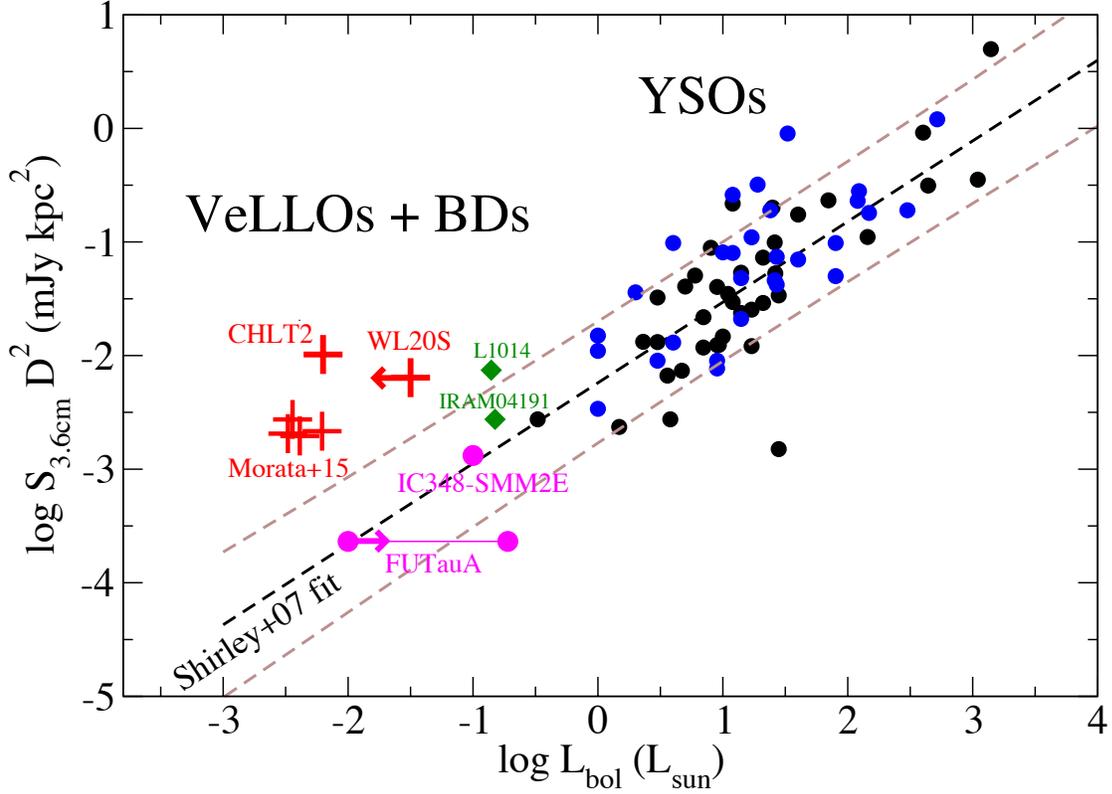}
\vskip-0.0cm
\caption{\small Centimeter luminosity at 3.6 cm vs. bolometric luminosity after Morata et al. (2015). Blue and black dots correspond to the data compiled by Anglada (1995) and Furuya et al. (2003), respectively, showing the relation 
for young stellar objects. Red plus signs correspond to the proto-BD candidates driving radio jets presented in Morata et al. (2015) and to the two candidate brown dwarfs discussed 
here. Very Low Luminosity Objects (VeLLOs) with detected 3.6 cm emission (Andr\'e et al. 1999; Shirley et al. 2007) 
are shown as green squares, and the big magenta circles correspond to the 3.0 cm observations of FU Tau A (this paper) and the 3.3 cm
observations of IC348-SMM2E (Palau et al. 2014, Forbrich et al. 2015). The black dashed-line is the fit performed by Shirley et al. (2007) 
to the young stellar objects, and the two brown dashed hyperbolae indicate the 1-$\sigma$ confidence band of the fit obtained by Shirley et al. (2007). The confidence band
was calculated following Weisberg (2013).}
\label{fig7}
\end{figure}

\pagebreak

\begin{deluxetable}{l c c c c c c c}
\tabletypesize{\scriptsize}
\tablecaption{Fields observed with the JVLA}
\tablehead{                        
\colhead{}                        &
\multicolumn{2}{c}{Phase Center} &
\colhead{}                 &
\colhead{}                              &
\colhead{}                              &
\colhead{}                              &
\colhead{}      \\
\colhead{}   &
\colhead{$\alpha_{2000}$}          &
\colhead{$\delta_{2000}$}           &
\colhead{Times} &
\colhead{Synthesized beam$^a$ }       &   
 \colhead{Rms$^b$} &                 
\colhead{Amplitude}  &
\colhead{Gain}   \\
\colhead{Source}                              &
\colhead{(h m s) }                     &
\colhead{($^\circ$ $^{\prime}$  $^{\prime\prime}$)}              &
\colhead{Observed} &
\colhead{($\theta_{maj} \times \theta_{min}$; ~PA)}  & 
\colhead{($\mu$Jy)} &
\colhead{Calibrator} &
\colhead{Calibrator} \
}
\startdata

FU Tau    & 04 23 35.40  & $+$25 03 03.1 &  5 & $2\rlap.{''}5\times2\rlap.{''}3$; $+$27$^\circ$  & 1.9 & J0542+4951  &  J0403+2600 \\
MHO 5  & 04 32 16.00 & $+$18 12 46.0 & 5 & $1\rlap.{''}8\times1\rlap.{''}6$; $+$40$^\circ$  & 3.8 & J0542+4951 & J0403+2600 \\
MMS 6-main & 05 35 23.48 & $-$05 01 32.3 & 5 &  $3\rlap.{''}3\times2\rlap.{''}2$; $-$15$^\circ$ &  3.0 & J0542+4951 & J0541$-$0541 \\
ISO-Oph 32  & 16 26 21.90 & $-$24 44 39.8 & 3 &  $5\rlap.{''}4\times2\rlap.{''}1$; $-$17$^\circ$ &  3.2 &  J1331+3030 & J1625$-$2527 \\
ISO-Oph 102 & 16 27 06.60 & $-$24 41 48.8 & 4 & $5\rlap.{''}1\times2\rlap.{''}4$; $-$5$^\circ$  & 3.0 & J1331+3030 & J1625$-$2527 \\
LS-RCrA 1 & 19 01 33.57 & $-$37 00 30.4 & 5 & $7\rlap.{''}0\times2\rlap.{''}2$; $+$10$^\circ$  & 3.2 & J0137+3309 & J1924$-$2914 \\
\enddata
\tablecomments{
                (a): From the images made concatenating all data. \\
                (b): At center of the images made concatenating all data. }
\end{deluxetable}

\pagebreak

\section{Comments on individual sources}

\subsection{FU Tau}

FU Tau is a young, wide brown dwarf binary system. Its components show a
projected angular separation of $5\rlap.{''}7$ (or 830 AU at a distance
of 145 pc, see below) with  a  position  angle  (PA)  of $\sim$145$^\circ$
(Luhman et al. 2009). The NW component, 
FU Tau A, has a spectral type of M7.25, corresponding
to a mass of 50 $M_{Jup}$, while the SE component, FU Tau B, 
has a spectral type of M9.25, corresponding to a mass of 15 $M_{Jup}$ (Luhman et al. 2009).
Monin et al. (2013) detected a molecular outflow associated with FU Tau, but given the angular resolution of 11$''$
of the IRAM 30 m telescope observations they could not establish which of the two components of the FU Tau binary
was producing it.  Since optical forbidden line emission, a reliable tracer of the shocks caused by outflow activity, has been detected in 
the spectrum of FU Tau A (Stelzer et al. 2010), Monin et al. (2013) assumed that this component is the driving source of the molecular outflow.

We adopted as the distance and proper motions of FU Tau the average values of the nearby stars V773 Tau (Torres et al. 2012) and HP Tau (Torres et al. 2009).
This gives an adopted distance of 145 pc and proper motions of $\mu_{\alpha}$cos$\delta$ = 15.5 mas yr$^{-1}$;
$\mu_{\delta}$ = $-$19.7 mas yr$^{-1}$ . In Figure 1 we show the radio continuum emission detected by us
with the positions of FU Tau A and B, after correction by these proper motions. As can be seen in the Figure the radio emission coincides with component
FU Tau A and this supports the proposition of Monin et al. (2013) that this is the source driving the molecular outflow.

\subsection{MHO 5}

This very low mass star was detected by Brice\~no et al. (1998). It has a spectral type of  M6.2.
Its estimated mass is 90 $M_{Jup}$ (Muzerolle et al. 2003), just above the hydrogen-burning limit.
Phan-Bao et al. (2011) reported an associated  bipolar CO outflow with
an estimated outflow mass of 7.0$\times$10$^{-5}$ $M_\odot$ and a mass-loss rate of 9.0$\times$10$^{-10}$ $M_\odot$ yr$^{-1}$.

We did not detect a radio source associated with MHO 5 at the 3-$\sigma$ level of 12 $\mu$Jy. Of the five sources detected in the field, the only one with a counterpart is
associated with 2MASS J04322946+1814002. This radio source is moderately time variable (showing a maximum-to-minimum flux density ratio of 1.6 
along the five epochs of observation)
and very bright with an average flux density of $\sim$67 mJy.
Based on its infrared properties, Gutermuth et al. (2009) classified it as a Class II young stellar object. However, Dzib et al. (2015) found no evidence of
proper motions, a result that favors an extragalactic nature for the source. In Figure 2 we present its radio continuum spectrum that shows a flux density rising with 
frequency and is consistent with an optically-thick synchrotron source. We tentatively propose that this source is an extragalactic high frequency peaker (HFP; Dallacasa et al. 2000).
Similar sources have been found in other regions of star formation (e.g. Rodr\'\i guez et al. 2014a; Dzib et al. 2015).

\subsection{MMS 6-main}

MMS 6-main is the brightest and the most compact submillimeter
continuum  source  in  the  Orion MMS  6  region  (Takahashi  et  al.
2009).   A compact molecular outflow lobe ($\sim$1000 AU) associated
with MMS 6-main and having
an estimated outflow mass of 3.3$\times$10$^{-4}$ $M_\odot$ and a mass-loss rate of 4.4$\times$10$^{-6}$ $M_\odot$ yr$^{-1}$ was reported by Takahashi \& Ho (2012).

The mm sources MMS 6-main and MMS 6-NE (Takahashi et al. 2009) coincide with two of the radio sources detected by us (see Figure 3).
MMS 6-main was detected previously at 3.6-cm with a very similar flux density (0.15 mJy; Reipurth et al. 1999) than in our detection.
To our knowledge, this is the first centimeter detection of MMS 6-NE.
This is an extensively studied region and of the 13 radio sources detected in the primary beam, 11 have previously reported counterparts (see Table 2).

There are two brown dwarfs in the field observed: TKK 755 and 2MASS J05351294-0502086, with spectral types of M7.75 and
M6.5 (Peterson et a. 2008), respectively. We did not detect associated radio sources with them at a 3-$\sigma$ upper limit of 16 $\mu$Jy.

\subsection{ISO-Oph 32}

This star, also known as  2MASS J16262189-2444397, has a spectral type of M6.5 (Manara et al. 2015).
This corresponds to a mass of $\sim$70 $M_{Jup}$, at the hydrogen-burning limit.
We did not detect an associated source at the 3-$\sigma$ level of 10 $\mu$Jy.

\subsection{ISO-Oph 102}

Our main goal in this field was the brown dwarf ISO-Oph 102. This object has a spectral type of M6 and a mass of 60 $M_{Jup}$
(Ricci et al. 2012). We did not detect an associated radio source at a 3-$\sigma$ upper limit of 9 $\mu$Jy.

However, in the field we detected two interesting sources. The first is WL 20S, a brown dwarf candidate (Alves de Oliveira et al. 2010)  that was detected by us with a flux density of
326$\pm$55 $\mu$Jy (Figure 4). This source was previously detected with the VLA with flux densities of 220$\pm$16 $\mu$Jy and 236$\pm$28 $\mu$Jy at 4.5 and 7.5 GHz, respectively (Dzib et al. 2013).
As discussed by Dzib et al. (2013) the radio source is associated with component WL 20S and not with one of the other two more massive components of the triple system, WL 20E and WL 20W
(see Fig. 4). The 2.7 mm dust emission is associated with WL 20S and WL 20W (Barsony et al. 2002).

The other interesting radio source detected in the field is associated with SR 12 (Struve \& Rudkjobing 1949). This is a T Tau binary (Simon et al. 1987) with K4 and M2.5 components (Gras-Vel{\'a}zquez \& Ray 2005), that
are separated by $0\rlap.{''}21$ (Kuzuhara et al. 2011). This angular separation corresponds to a physical separation of 29 AU at a distance of 137 pc (Ortiz-Le\'on et al. 2016).
This T Tau binary, also known as SR 21AB, has associated the exoplanet candidate SR 12C, with an estimated mass of $\sim$13 $M_{Jup}$ (Kuzuhara et al. 2011).
SR 12C appears projected by $\sim 8\rlap.{''}7$ ($\sim$1200 AU at a distance of 137 pc) to the south of SR 12AB and is the widest-separation substellar companion candidate to a T Tau binary known (Kuzuhara et al. 2011).

The radio source was previously detected with the VLA with flux densities of 160$\pm$37 $\mu$Jy and 87$\pm$12 $\mu$Jy at 4.5 and 7.5 GHz, respectively (Dzib et al. 2013). There are no reported optical or infrared 
individual positions for the stars in this binary system and we could not determine with which of the stars is the radio emission associated (Figure 5). The extra-solar planet is not detected at
a 3-$\sigma$ upper limit of 24 $\mu$Jy. This relatively large noise is due to the primary beam correction applied to this region that is far ($\sim2\rlap.'5$) from the phase center.

\subsection{LS-RCrA 1}

Our main goal in this field was the brown dwarf 2MASS J19013357-3700304, with a spectral type M6.5 (Fern\'andez \& Comer\'on 2001). We did not detect an associated radio source at a 3-$\sigma$ upper limit of 10 $\mu$Jy.
In the field we detected the source CHLT 2 (see Fig. 6), a brown dwarf candidate (Feigelson et al. 1998).
This source was first detected at radio wavelengths by Brown (1987) and is sometimes referred to as Brown 5.
Since then, it has been detected in several radio studies of the core of the Corona Australis molecular cloud (e.g. Suters et al. 1996; Forbrich et al. 2006; Choi et al. 2008; Miettinen et al. 2008; Liu et al. 2014).
The radio source is time variable, with centimeter flux densities ranging from 0.2 mJy (Forbrich et al. 2006) to 4.4 mJy (Suters et al. 1996). However, it did not show evidence of
variability in the five epochs we observed it. From measurements at 3.5 and 6.2 cm, Choi et al. (2008) determined a spectral index of $-$1.3$\pm$0.3 for year 1996
and of +1.06$\pm$0.13 for year 2005. A highly negative spectral index rules out a free-free nature for the emission (Rodr\'\i guez et al. 1993). 
For observations made in 1986 and 1992, Suters et al. (1996) determined spectral indices in the range
of $-$0.28$\pm$0.30 to +1.05$\pm$0.17. At least in one epoch the source showed evidence of circular polarization (Choi et al. 2009). 

CHLT 2 is associated with a faint (K = 16.4) near-infrared source
that would be a brown dwarf if it lies in the CrA cloud and is $<$10$^7$ yr old (Feigelson et al. 1998). Miettinen et al. (2008) note that their
radio properties, in particular its persistent strong emission, do not support the brown dwarf classification. Liu et al. (2014) propose that this source is a radio galaxy.

\section{Discussion}

There are twelve young brown dwarfs and four young brown dwarf candidates in the six fields imaged.
We detected only one brown dwarf and two brown dwarf candidates. The small number of detections precludes a firm statistical conclusion,
but this result suggests that brown dwarf candidates have larger radio flux densities than the brown dwarfs.

Morata el al. (2015) have recently discussed the expected radio luminosity for young brown dwarfs. Extrapolating the well-known correlation between radio luminosity and bolometric luminosity
(their Figure 5) one estimates that for the brown dwarf region the expected luminosity is roughly bracketed by $S_{3.6~cm} \simeq 10^{-4.0\pm0.5}$ mJy kpc$^2$.
Assuming that the 3.6 cm flux density is similar to the 3.0 cm flux density measured by us and since FU Tau A is at a distance of 145 pc, we find that for this source
$S_{3.6~cm} \simeq 10^{-3.7}$ mJy kpc$^2$, in the range of the equation given above. A similar agreement is obtained for IC 348-SMM2E (Palau et al. 2014; see below).

To see this more clearly, we have updated the figure of Morata et al. (2015)
showing the relation between radio luminosity and bolometric luminosity with our new observations (Figure 7). To do the figure, we have assumed for FU Tau A a range of bolometric luminosities from 0.19 L$_\odot$ (Luhman et al. 2009) to 0.01 L$_\odot$. This last luminosity is estimated by using the relation between Spectral Type (ST) and effective temperature given by Baraffe et al. (2015). 
For a ST M7.25 for FU Tau A (Luhman et al. 2009) the corresponding effective temperature is 2700 K. 
We then converted the effective temperature to luminosity by using the relation of Luhman et al. (2003, their Fig. 8). It should be noted, however, that
since FU Tau A is overluminous for its effective temperature (Scholz et al. 2012b), this lower limit is uncertain. For the case of IC348-SMM2E, we used the bolometric luminosity estimated by Palau et al. (2014), and the new deep 3.3 cm observations reported in Forbrich et al. (2015), where the source was detected for the first time at this wavelength. We note that Forbrich et al. (2015)
estimate a spectral index of 0.4 for IC348-SMM2E, suggesting a thermal radio jet. The results of Rodr\'\i guez et al. (2014b) are also consistent with
a positive spectral index. Figure 7 shows that the confirmed brown dwarfs FU Tau A and IC348-SMM2E follow very well the well-known relation 
between radio luminosity and bolometric luminosity, suggesting that this relation extends down to the brown dwarf regime and that at least these two brown dwarfs seem to form as a scaled-down version of low-mass stars.
Indeed, FU Tau has previously been proposed to be an example of star-like formation, based on its isolated position, its wide companion, its disk, and its outflow (e.g. Luhman et al. 2009; Stelzer et al. 2010).

In contrast, the radio luminosities of the sources associated with the candidate brown dwarfs are much larger than expected. In the case of WL20 S (assuming a distance of 140 pc; Ortiz-Le\'on 2016)
we obtain $S_{3.6~cm} \simeq 10^{-2.2}$ mJy kpc$^2$,
while for CHLT 2 (assuming a distance of 130 pc; Neuh{\"a}user \& Forbrich 2008) we obtain $S_{3.6~cm} \simeq 10^{-2.0}$ mJy kpc$^2$.
These radio luminosities are more than an order of magnitude larger than expected. We show this more clearly in Figure 7, where we 
used the same strategy as in the case of FU Tau A to estimate the bolometric luminosity of WL20S, for which we assumed that the Spectral Type should be M6.5 or later, while for CHLT2 the bolometric luminosity is estimated to 
be 0.006 $L_\odot$, assuming a spectral type M9 (adopted by Miettinen et al. 2008). Taking into account that the radio luminosity-bolometric luminosity relation seems to hold for the confirmed brown dwarfs discussed above, 
our results for WL20 S and CHLT 2 suggest that these two candidates might be instead background sources of different nature.

%Discutir si FU Tau A cae en la correlacion luminosidad bolometria-luminosidad de radio.

%Discutir si el relativamente alto flujo indica que las brown dwarf candidates son otra cosa.

%?Poner una tabla con las brown dwarfs y candidate brown dwarfs detectadas o con limites superiores?

%?buscar contrapertes de  CHLT 2 en 2MASS, Herschel o algo asi?

\section{Conclusions} 

The high sensitivity of the Jansky VLA allows the detection of faint, previously unreported sources in
regions of star formation. The main results of our study
can be summarized as follows.

1. We observed six regions with young brown dwarfs associated with outflows. We detected a total of 49 compact sources, of which 24 are new detections.

2. The only \sl bona fide\rm ~brown dwarf detected by us is FU Tau A. Its radio luminosity is consistent with the extrapolation of the radio luminosity-bolometric luminosity correlation down to the substellar regime. Assuming that the centimeter emission from FU Tau A comes from a faint thermal jet, our findings indicate that the radio luminosity-bolometric luminosity correlation seems to extend to the brown dwarf regime, supporting the view that brown dwarfs form as a scaled-down version of low-mass stars.

3. We detected radio sources in association with two brown dwarf candidates (WL 20S and CHLT 2) in the regions studied. Their radio luminosities are more than an order of magnitude larger than expected from the radio luminosity-bolometric luminosity correlation, questioning the brown dwarf classification.

\acknowledgments

We thank an anonymous referee for valuable comments that improved the paper.
This research has made use of the SIMBAD database,
operated at CDS, Strasbourg, France.
LFR, LAZ and AP are grateful to CONACyT, Mexico and DGAPA, UNAM for their financial
support.

\begin{deluxetable}{l c c c c c c  c}
\tabletypesize{\scriptsize}
\tablecaption{Sources detected}
\tablehead{                        
\colhead{}                        &
\colhead{}                        &
\multicolumn{2}{c}{Position} &
\colhead{}                              &
\colhead{}                              & \\
\colhead{}   &
\colhead{}   &
\colhead{$\alpha_{2000}$}          &
\colhead{$\delta_{2000}$}           &
\colhead{Flux Density$^a$ }       &                            
\colhead{}  & \\
\colhead{Field}                              &
\colhead{Counterpart}                              &
\colhead{(h m s) }                     &
\colhead{($^\circ$ $^{\prime}$  $^{\prime\prime}$)}              &
\colhead{($\mu$Jy)}  & 
\colhead{Notes$^b$}  &
}
\startdata
FU Tau &  -- & 04 23 27.62 & $+$25 06 08.8 &448$\pm$25 & \\
FU Tau  &   --  &  04 23 28.87 & $+$25 02 40.0 & 24$\pm$4 &  \\
FU Tau &   -- &  04 23 31.84 & $+$25 02 08.1 & 46$\pm$2 &  \\
FU Tau & FU Tau A &  04 23 35.38 & $+$25 03 02.4 & 11$\pm$ 2 &  BD \\
FU Tau &  -- & 04 23 35.88 & $+$25 04 34.3 & 161$\pm$10 &  \\
FU Tau & -- & 04 23 36.09 & $+$25 04 27.7 & 199$\pm$9 &  \\
FU Tau & -- & 04 23 36.57 & $+$25 00 52.8 & 233$\pm$7 &  \\
FU Tau & -- & 04 23 39.62 & $+$25 04 30.4 & 21$\pm$3 &  \\
FU Tau & -- & 04 23 40.75 & $+$25 04 10.6 & 30$\pm$2 &  \\
FU Tau & -- & 04 23 41.44 & $+$25 04 25.3 &46$\pm$7 &  \\
FU Tau & -- & 04 23 48.81 & $+$25 05 26.8 & 775$\pm$43 & \\
  & & & & & & & \\
MHO 5  & -- &   04 32 05.71 & $+$18 13 02.5  & 155$\pm$10 & \\
MHO 5  & -- &   04 32 17.24 & $+$18 14 25.2 & 49$\pm$6 & \\
MHO 5  & -- &   04 32 17.45 & $+$18 14 22.0 & 663$\pm$14 & Time Variable(2.2) \\
MHO 5  & -- &  04 32 17.64 & $+$18 10 06.9 & 403$\pm$33 & \\
MHO 5  & 2MASS J04322946+1814002 &  04 32 29.46 & $+$18 14 00.2 & 67100$\pm$1200 & Time Variable(1.6)  \\
    & & & & & & & \\
MMS 6-main & V2282 Ori &   05 35 16.16 & $-$05 00 02.7 & 63$\pm$8 & \\
MMS 6-main & [TKT2002b] IRS 3 &   05 35 18.28 & $-$05 00 33.6 & 374$\pm$11 &  \\
MMS 6-main & V1733 Ori &   05 35 20.75 & $-$04 58 34.1 & 258$\pm$23 & Time Variable(3.7) \\
MMS 6-main & -- &  05 35 22.05 & $-$05 00 47.8 & 74$\pm$5 & Time Variable(1.5) \\
MMS 6-main & [THT2013] OMC3-SMM 6  &  05 35 22.48 & $-$05 01 14.5 & 88$\pm$6 & Time Variable(1.9) \\
MMS 6-main & [TSO2008] 14, MMS 6-main &  05 35 23.43 & $-$05 01 30.4 & 140$\pm$10 & \\
MMS 6-main & [THT2009] IRS 3, MMS 6-NE &  05 35 23.52 & $-$05 01 29.1 & 128$\pm$20 & \\
MMS 6-main & 2MASS J05352431-0501204 &  05 35 24.34 & $-$05 01 20.7 & 52$\pm$3 &  \\
MMS 6-main & HOPS 84 &  05 35 26.56 & $-$05 03 55.2 & 363$\pm$17 & \\
MMS 6-main & HOPS 85 &  05 35 28.19 & $-$05 03 41.2 & 69$\pm$5 & \\
MMS 6-main & HOY J053529.50-045952.7 &  05 35 29.79 & $-$04 59 50.5 & 296$\pm$13 & \\
MMS 6-main & -- &  05 35 30.11 & $-$05 03 28.5 & 55$\pm$5 & \\
MMS 6-main & [RRC99] VLA 2 &  05 35 37.20 & $-$05 00 55.1 & 296$\pm$21 & \\
  & & & & & & & \\
ISO-Oph 32 & 2MASS J16262367-2443138 & 16 26 23.71 & $-$24 43 14.4 & 90$\pm$10 & \\
ISO-Oph 32 & -- & 16 26 24.45 & $-$24 45 24.8 & 109$\pm$9 & \\
ISO-Oph 32 & -- &  16 26 24.92 & $-$24 43 34.2 & 77$\pm$9 & \\
    & & & & & & & \\
ISO-Oph 102 & GBS-VLA J162657.84-244201.6 & 16 26 57.84 & $-$24 42 01.7 & 160$\pm$9 & Time Variable(2.2) \\
ISO-Oph 102 & -- & 16 27 02.15 & $-$24 38 44.6 & 277$\pm$11 & Time Variable(2.7) \\
ISO-Oph 102 & -- & 16 27 02.72 & $-$24 39 00.1 & 48$\pm$6 &  \\
ISO-Oph 102 & 2MASS J16270451-2442596 & 16 27 04.51 & $-$24 43 00.0 & 110$\pm$4 & Time Variable(2.0) \\
ISO-Oph 102 & 2MASS J16271117-2440466 &  16 27 11.16 & $-$24 40 46.7 & 76$\pm$9 & Time Variable(1.6) \\
ISO-Oph 102 & LFAM 29  &  16 27 14.67 & $-$24 39 20.6 & 153$\pm$24 & \\
ISO-Oph 102 & WL 20S & 16 27 15.70 & $-$24 38 45.4 & 326$\pm$55 & BDc \\
ISO-Oph 102 & EM* SR 12 &  16 27 19.50 & $-$24 41 41.0 & 103$\pm$18 & \\
  & & & & & & & \\
LS-RCrA 1 & [B87] 1 & 19 01 18.26 & $-$37 00 02.4 & 621$\pm$22 & \\
LS-RCrA 1 & -- & 19 01 18.82 & $-$37 02 53.5 & 1180$\pm$200 & Time Variable(1.7) \\
LS-RCrA 1  & -- &  19 01 23.30 & $-$37 02 15.5 & 200$\pm$15 &  \\
LS-RCrA 1   & -- &  19 01 26.44 & $-$36 59 22.2 & 79$\pm$8 &  \\
LS-RCrA 1   & 2MASS J19012717-3659085 &  19 01 27.22 & $-$36 59 08.3 & 47$\pm$3 &  \\
LS-RCrA 1   & -- &  19 01 34.57 & $-$37 03 02.9 & 1740$\pm$150  & Extended ($\sim$6'') \\
LS-RCrA 1   & V709 CrA  &  19 01 34.89 & $-$37 00 56.7 & 211$\pm$7 & Time Variable(3.3) \\
LS-RCrA 1   & 2MASS J19014156-3658312  &   19 01 41.61 & $-$36 58 31.0 & 564$\pm$19 & Time Variable(1.6)  \\
LS-RCrA 1   & CHLT 2 &  19 01 43.31 & $-$36 59 11.7 & 614$\pm$14 & BDc  \\
LS-RCrA 1  & -- &  19 01 44.36 & $-$37 01 54.7 & 92$\pm$12 &  \\
\enddata
\tablecomments{
                (a): Total flux density corrected for primary beam response,
obtained from the images made concatenating all data.
(b): BD = brown dwarf, BDc = brown dwarf candidate.}
\end{deluxetable}

\clearpage

\end{document}